# Excited-state dynamics of structurally characterized crystal of $Sn_xSb_{1-x}$

Prince Sharma[1, 2], M.M. Sharma[1,2], Kapil Kumar[1, 2], Mahesh Kumar[1, 2], V.P.S. Awana[1, 2*]

[1]CSIR-National Physical Laboratory, Dr. K.S. Krishnan Marg, New Delhi-110012, India
[2]Academy of Scientific and Innovative Research (AcSIR), Ghaziabad, 201002, India

## ABSTRACT

The topological behaviour of heavy metal alloys opens a vast area for incredible research and future technology. Here, we extend our previous report about the superconducting properties of $Sn_{0.4}Sb_{0.6}$ along with the compositional variation of Sn and Sb in $Sn_xSb_{1-x}$ (with (X=0.5 and 0.6)) to study the detailed optical properties. Structural and morphological details of grown crystal are carried from the previous study. Further, the samples are excited by a pump of 2.61 eV with a broad probe of 0.77-1.54 eV in the NIR regime for transient reflectance ultrafast studies (TRUS) measurements. The differential reflectance profile shows an unprecedented negative magnitude, and the average power-dependent analysis of this negative trend has been analyzed. This article not only provides evidence of band filling phenomenon in the samples but also shows that with the variation of average power, there is a definite increase in the excited charge carriers, and thereby enhancing the band filling response. The estimated value of the bandgap between the band filled states and valence state is also determined from these studies. The nonlinear properties and bandgap analysis of the studied topological alloys and similar materials help in the advancement of various nonlinear optical applications.

[*]**Corresponding Author**

Dr. V. P. S. Awana: E-mail: awana@nplindia.org
Ph. +91-11-45609357, Fax-+91-11-45609310
Homepage: awanavps.webs.com

# INTRODUCTION

Topologically protected states in a material give rise to a set of symmetries[1–6]. The incredible trend of the surface charges due to these regularities has gained tremendous attention. Recently, heavy metal alloys too showed these kinds of characteristics due to the presence of strong spin-orbit coupling[1–10]. It makes them a novelistic candidate in the topological phase. The categories of materials that include them are the topological insulators (TIs) and topological semimetals (TSM)[1–4, 8–13]. The binary alloys made up of Bismuth (Bi), and TSM Antimony (Sb) is one of the artless examples of fusion exhibiting the Dirac cone nature. More precisely, $Bi_xSb_{1-x}$ is the first material whose eccentric characteristics resemble the three-dimensional topological insulator[10, 12, 14]. A particular stoichiometric ratio of Bi and Sb causes band inversion as x falls in between 0.07 and 0.22[10, 12, 14]. The specific range of 'x' opens a bandgap due to the band reversal and makes the surface conducting. TI shows numerous applications in the field of thermoelectric, spintronics, superconducting as well as optoelectronics applications. TI sub-category topological superconductor (TSC) has potential use in fault-tolerant quantum computing too[12, 15–17] as well as terahertz generation[18, 19]. Recently, doping of TI with a suitable dopant and optimizing the stoichiometry of doping with its physical properties have opened up wide research avenues. The Selenide and Telluride of Bismuthdo not show the superconducting behavior without the dopant atoms such as $A_xBi_2Se_3$, where A= Cu, Sr, Nb, and TI. It is one of the examples of such kind of modification in the usual TI to form TSC. There are a few more topological materials that show superconductivity in their intrinsic form or after doping, such as $TI_5Te_3$, $Sn_{1-x}In_xTe$, $PbTaSe_2$, $Au_2Pb$, and $PbTaSe_2$[20–25]. Therefore, the wide compositional tunability of alloys such as $Bi_xSb_{1-x}$ alloys shows the desired verdict for such TSC. The phase diagram and the physical properties of another binary alloy system, i.e., $Sn_xSb_{1-x}$ has also been studied recently[26–29]. It shows three single-phase regions with compositional stoichiometry. Regardless of the physical properties, it is more prominent to investigate the excited charge carrier's dynamics of such a unique system. The structure of SnSb consists of stacking of alternate layers of Sn and Sb, such as Sn-Sb-Sn-Sb-Sn-Sb-Sn, where Sb is stacked along the c axis[26–29]. The Sn-Sb system has come into the picture due to its electrochemical performance. But its physical properties also signifies its superconducting nature where it is reported that it is a type II superconductor below 1.5 K at x=0.5[26–29]. Recently, in our previous report, there is an enhancement in the superconducting critical temperature at x=0.4. The $Sn_{0.4}Sb_{0.6}$ shows the superconducting nature below 4K [30]. But, beyond these valuable properties, it has not yet been investigated for its carrier dynamics. In this article, the $Sn_xSb_{1-x}$ TI alloys are used to investigate its excited state carrier dynamics where the composition is varied as x = 0.4, 0.5, and 0.6. We deal with the band related phenomenon, which includes the band filling and renormalization in the sample. Further, its carrier relaxation lifetime is used to find the approximate bandgap of the system. This study is prolific in developing the TI alloys for nonlinear optics and optoelectronics applications.

**EXPERIMENTAL DETAILS**

$Sn_xSb_{1-x}$ TI crystal is grown by a solid-state reaction route by melting a stoichiometric mixture of high purity powders of Sn (>4N) and Sb (>4N) in a simple automated muffle furnace at a temperature of $900^0C$ for 48 hours, and the mixture was vacuum-sealed quartz ampoule. Then this melt is allowed to cool slowly at $2^0C/h$. up to $300^0C$ and annealed at this temperature for 72 hours. This annealed sample is directly quenched in ice water to avoid the formation of any low-temperature phase. The heat treatment is the same for all the samples as shown in S.Fig-1(a) (supplementary figure 1). The samples that come out through this heat treatment are a shiny silver color. The $Sn_xSb_{1-x}$ sample has been characterized by using various techniques such as PXRD (Powder X-ray Diffractometry), SEM (Scanning Tunneling Microscopy), and EDAX (Energy-dispersive X-ray Spectroscopy). PXRD has been performed using Rigaku made SmartLab X-ray diffractometer having $CuK_α$ radiation of 1.5418Å wavelength of X rays. The composition and visualization of the morphology of as-grown $Sn_xSb_{1-x}$ crystal have been carried out using EDAX and SEM with (Zeiss EVO-50), respectively. The magnetic analysis using QD-PPMS (Physical Property Measurement System) has also been carried out of $Sn_{0.4}Sb_{0.6}$. The Helios ultrafast spectroscopy system (from ultrafast systems) in the reflectance geometry has been used for evaluating the excited state carrier dynamics of the flakes of $Sn_xSb_{1-x}$ crystal. The whole spectroscopy system consists of a mode-locked laser, an amplifier, TOPAS (Optical parametric amplifier), delay stage, and spectrometer[18, 31, 32]. Mode-locked laser (Micra form Coherent) produces a laser pulse of ~35 fs with an average power of ~280-400 mW. An amplifier (from Coherent) amplifies the mode-locked laser to 3W-4 W average power with the help of Ti-Sapphire crystal. The amplified laser beam splits into 70-30 as 70% part is used as a pump beam while the 30% used as a probe beam through the delay stage. The 800 nm amplified laser is passed through TOPAS for wavelength selection. A broad range of pumps can be used using TOPAS, and the sapphire crystals are used to generate a broad range of probe beam in the entire range of wavelengths inside the spectrometer. It includes from 320 to 1600 nm in the spectra. The Helios spectrum is calibrated by the reference material of ZnTPP (soon to be released as BND$^{TM}$) in NPL.

**RESULT AND DISCUSSION**

The $Sn_xSb_{1-x}$ where, x=0.4, 0.5, and 0.6 have been synthesized through a conventional solid-state reaction route. Figure 1 shows the Rietveld refinement of the PXRD data of $Sn_{0.4}Sb_{0.6}$. This refinement shows that the sample is crystallized in a rhombohedral structure with an R -3 m space group. Also, it shows the presence of two phases of the same space group with different c parameters, as reported earlier by our group. In $Sn_{1-x}Sb_x$, both Sn and Sb occupy (0,0,0) atomic positions. Details of refinements and lattice parameters can be found in reference[30]. The presence of two unit cells results in the structure of $Sn_{1-x}Sb_x$ to contain alternate layers of Sn and Sb, as shown in the inset of Fig.1 details that can be found in reference[30]. The $Sn_{0.4}Sb_{0.6}$ has a layered structure, as shown in the right inset of figure 1. It shows the SEM image of the crystal at a resolution of 2 µm. This layered morphology of the crystal confirms the alternative stacking of the rhombohedral Sn and Sb lattice as predicted in earlier results[26–30]. There are few defects in the crystal, as depicted in the SEM image. The remaining inset in figure 1 shows the M-H curve of the alloys at a range of low temperatures from 2 K to 3.7 K. It confirms the existence of bulk type-II superconductivity at 3.7 K as shown from the wide opening of the M-H curve[30]. The superconductivity in this sample is due to the topological character of the same which is further confirmed through the density functional theory (DFT) as shown in S.Fig.2. The atoms position and the lattice parameters are considered from the refinement of the XRD pattern. The K points, as well as the stable path, are used which are calculated from the SeeK-path: the *k*-pathfinder and visualizer[33]. The DFT (Density Functional Theory) calculations confirm the topological phase of the system as shown in S.Fig.2. Comparing the bands, one with SOC (Spin-orbit-coupling) and another without SOC, there is a clear indication of band inversion. This inversion induces a gap between the valence band and conduction band[29]. Thus, it confirms that the SOC is effective in the case of SnSb alloys. Figure 2 shows the comparison between the XRD pattern of all the three crystalline samples. It is confirmed from the XRD pattern that there is a slight shift in the peaks among different compositions. There are small peaks in the case of $Sn_{0.4}Sb_{0.6}$ and $Sn_{0.5}Sb_{0.5}$ which confirms the secondary phase in both the samples as suggested in one of the samples in the previous report[30]. The inset of figure 2 shows the layered structure of $Sn_{0.5}Sb_{0.5}$, and $Sn_{0.6}Sb_{0.4}$ respectively. The EDAX analysis is shown in S.Figure 1(b) of all the SnSb alloys.

After evaluating the crystallite and the superconducting behavior of the crystal, it is studied by TRUS. A small flake is taken out from $Sn_{0.4}Sb_{0.6}$ and used for the studies of excited-state dynamics. The sample has been pumped with a 475 nm (2.61 eV). An average power-dependent study has been carried out in which 0.5, 1.0, and 1.5 mW average laser power of 2.61 eV is used. A broad range of probe beam has been used in the NIR regime, which covers from 800-1600 nm (0.77-1.54 eV) at all the considered average powers. Figure 3 shows the 3D pictorial response of the flake at three different average powers as starting from lower to high as we go from A to C, respectively. The different axis represents a different parameter as the Z-axis epitomize the differential reflectance response as the colour represents the high and low reflectance by the sample. At the same time, the X and Y-axis spectacle the probe wavelength and the time in logarithmic scales measured in ns, correspondingly. The broad probe wavelength shows a fascinating aspect in the low wavelength range that is from 800-1200 nm. There is an increase in the differential reflection signal with average power. The physical reason for this upsurge in differential reflectance is the band filling phenomenon[31, 34–38]. It resembles that after photoexcitation, the excited population of carriers is accumulated in the allowed optical transitions. The rise of reflectance is in direct accordance with the excited carriers density in this permissible optical states[37–39]. The increase of excited carriers at these low power is due to the band filling effect.

The excited particles are in quasi-equilibrium, which occupies the available states from the lowest states due to the principle of energy minimization. It results in the filling of available optical states near the bottom of the conduction band by electrons and holes in the top of the valence band, which is called band filling. Hence, due to the band filling effect, the absorption coefficient decreases at energies higher than the bandgap. Similarly, by using the Kramers-Kroning integrals, it can be shown that the effective refractive index decreases for below and near bandgap, and the change in refractive index is positive for the above bandgap energies. Hence, on increasing the average power of pumped energy, there is a decrease in the nonlinear absorption coefficient and refractive index near bandgap due to the band filling phenomenon. The differential reflectance shows a fascinating aspect, too. It is evident that with increasing the average power of the pumped laser, there is an increase in the magnitude of differential reflectance due to absorption bleaching[31, 34, 36–38, 40, 41]. The escalation of the signal is due to the generation of carriers that occupies the optically-coupled states. At frequencies higher than the probing ones, the real part of the complex refractive index increases, and thereby there is an increase in reflectance. The 3-D signal has then investigated for the dependence of the differential reflectance with the probe wavelength and its delay individually in order to get the specific relevance of the physical process as described earlier and to know the carrier's lifetime. Thereby, it predicts the average power dependence on the charge carriers also.

Figure 4 shows the differential reflectance response concerning probe wavelength and delay, respectively, in 4(a) and 4(b). The sample has been excited with the pump energy of 2.61 eV, an average power of 0.5 mW. Figure 4(a) speculates the variation of reflectance after excitation by the laser pulse in the broad NIR regime. An overall view of the response is demonstrated in a color map, as shown in figure 3. The data has chirped to have probe delay correction as confirms from the no signal at 0 s as in figure 4(a). There is a broad reflectance in the entire NIR regime of reflectance. It is expected to have this kind of signal as to when the excited photo carriers occupy probed states; there is a reduction in reflectance. As the probe is delayed with respect to the pump arrival, different features in the differential reflectance are observed and can be distinguished quite easily. The probe wavelength is divided into two halves, one from 800-1200 nm and another from 1200-1600 nm termed as F and H discretely. There is an increase in reflectance (that is more negative absorbance) as we go from 250 fs to 500 fs in the H region, while in the F region, there is a decrease is in the negative reflectance. The maximum reflectance has been shown at this decay time. At 750 fs, there is a decrease of negative reflectance, too, in the H region. The trend continues for both the region as it approaches zero at 1 ps. The positive reflectance is first observed at 2.5 ps in the F regime, and it maximizes at the delay time of 10 ps. The H region shows less reflectance as approximately close to no signal. The response exists up to 50 ps and finally goes to zero at 3 ns. The unique trend resembles the bleaching of photoexcited carriers at a low decay time range of 500 fs due to the pumped laser pulse. Secondly, the positive reflectance from 2.5ps to 50 ps predicts the band filling effect as the electrons fill from lower conduction band energies to higher levels. It shows how a bleached system goes into the band filling state with time.

The kinetic decay profile from the differential reflectance signal is shown in figure 4(b) for the 0.5 mW average power. It is in accordance with the differential reflectance signal in terms of probe wavelength. The kinetic decay profile is considered at four different probe wavelengths such as 900 nm, 1100 nm, 1300 nm, and 1500 nm to predict the carrier lifetime at all these respective energies. Comparing the negative rise of the decay among these probes, it shows that with an increase in the wavelength, there is an increase in the negative differential signal. It predicts that at a higher wavelength that is at lower energy levels, the number of excited charge carriers is higher than that at higher energies due to the bleaching effect. The effect dominates in the lower energies than the higher ones as most of the carrier accumulates at these energy levels. Another unique difference among probe delayed kinetics is the relaxation of the excited carriers. Figure 4(b) shows no positive differential reflectance signal at 1100 nm, 1300 nm, and 1500 nm case. While at 900 nm, the decay signal goes positive through the relaxation in picosecond time regimes. This constructive decay signal again resembles the concept of band filling in the higher energy regime as anticipated above with the help of 3-D color signal.

The kinetics are fitted using the Surface Xplorer software (from Ultrafast Systems) by considering the decay equation with two carriers lifetimes $t_1$ and $t_2$[31, 32]. The $t_1$ is the rise or trapping time, and $t_2$ is the decay time or the recombination time of the charge carriers. Figure 4(b) shows the kinetic decay fitting in the nanoseconds regime while the inset of figure 4(b) shows the relaxation profile up to a few femtoseconds. The values of the fitting parameters predict that the bleaching time is in few femtoseconds, while the band filling persists for long picoseconds. The trend confirms that the charge carriers are excited from its ground state to higher states by the femtoseconds laser pulse within a small time scale. While the substantial picoseconds time of band filling phenomenon too confirms from the fitting parameters as shown in the S.Table.I (supplementary information table). Hence, the bleaching of charge carriers initially due to laser pulse causes band filling in a higher time scale. The amplitude of the fitting time constants depicts about the carriers in percentage and gives substantial information at the concerning probe wavelengths. Most of the kinetic signal fits with the trapping time signal as approximately above 90%, while the recombination time's amplitude is quite low. Most of the carriers that are excited from its ground state to the excited state and are relaxed through interband relaxation while very few go to the band filling state as it is on the much higher energy level.

The crystal response at 1.0 mW and 1.5 mW average power is studied and is shown in the S.Figures 3 and 4. A similar trend of the differential reflectance has been observed in these spectra, as in the case of 0.5 mW. But, there is an increase in the reflectance amplitude, which resembles the upsurge in excited carriers with the pumped pulse power. The time-dependent trends of the differential reflectance have been observed. It also has a similar kind of fashion as in the case of 0.5 mW and confirms the concept of band filling and the bleaching effect in the $Sn_{0.4}Sb_{0.6}$ crystals even in high average power. Additional, there is an increase in the differential reflectance at the lower wavelength, which shows the enhancement in the band filling effect with the average power. The kinetics of these two average power has been fitted with the same fitting equation at the same four probes wavelengths as in the earlier case. The fastidiousness of fitting and the values of the appropriate parameter has been shown in S.Table II and III. The kinetic trends of maximum differential reflectance at the upper wavelength are also demonstrated at the higher average powers as the signal at 1500 nm has maximum reflectance as compared to other wavelengths. It confirms the bleaching of carriers among different energy levels is independent of the average power, and most of the excited charge accumulates to the 0.82 eV energy level in all the average power. While comparing the magnitude of decay kinetics among different average powers, there is a small increase in the amplitude of the signal from 0.5 to 1.5 mW. The trends show that with an increase in the average power, there is an upsurge in the excited carriers but not at the different bands in the sample.

The behavior of band filling at higher energies is also confirmed from the average power-dependent studies. The 900 nm kinetic signal shows a definite upsurge at the higher picoseconds time scale. It is following the conduction band filling phenomenon by excited electrons. But with increasing the average power, the 900 nm kinetic signal becomes more positive from 60 ps to 100 ps as compared to lower average power. This shows that with an increase in the average power, there is a definite rise of another kinetic probe wavelength at 1100 nm which shows a positive differential reflectance at this decay time range. It endorses increasing of the band filling phenomenon with the average power as predicted by the differential reflectance with probe wavelength. Figure 5(a) shows the t1 and t2 carriers lifetime variation with the average power at different probe wavelengths. Considering the inset which is showing the t1 variation, it depicts that with an increase in the average power, there is no such rise in the decay time. While with the probe wavelength, there is an increase in the t1 as we go from higher energy states to lower from 1.55 eV to 0.77 eV. The t2 is the slow lifetime component of the carrier as it results from the cooling of the hot carriers to the lattice temperature at equilibrium. The trend of t2 with the average power is quite insignificant as it changes randomly with the probe wavelength. At the same time, it decreases with the average power, which resembles the small persistence of band filling phenomena at higher average power. The trend of the t1 and t2 has helped to estimate a particular confined range of the bandgap posed by the crystal concerning the probe wavelength too[36, 40, 41]. Figure 5(b) shows the variation of t1 and t2 of all the three average power with reference to the probe wavelength. The t1 shows almost a linear relation with the probe wavelength regime, while t2 shows a peak in the value at around 1301 nm. In order to find the exact peak in the graph, it has displayed by the short dot line. The peak that comes out by fitting the t2 data is 1301 nm. This peak corresponds to the bandgap of the sample and it basically shows the gap between the band in which band filling occurs and the valence band. The estimated bandgap that found out by this analysis is around 0.95 eV.

After evaluating the $Sn_{0.4}Sb_{0.6}$ with the different average power from 0.5 to 1.5 mW, the two other compositions of SnSb alloys are studied with the same pump energy as well as in the same range of probe wavelength. The 3D pictorial responses of the two different flakes from $Sn_{0.5}Sb_{0.5}$ and $Sn_{0.6}Sb_{0.4}$ at three different average power as taken above is shown in S.figure 5 and 6 respectively. These two crystal flakes also show the same trend as in the case of the prior one. There is also an increase in the differential reflection signal with average power. As mentioned above, the physical reason for this upsurge in differential reflectance is due to the band filling phenomenon[31, 34–38]. The rise of reflectance is in direct accordance with the excited carriers density in this permissible optical states[37–39]. The S.Figures 7, 8, and 9 show the differential reflectance response of $Sn_{0.5}Sb_{0.5}$ at concerning probe wavelengths and delays, with three different average power 0.5, 1.0, and 1.5 mW respectively. While the average power-dependent differential reflectance of $Sn_{0.6}Sb_{0.4}$ is as shown in S.Figures 10, 11, and 12 with their respective kinetics at four different probe wavelengths. Individually, both the crystals of $Sn_{0.5}Sb_{0.5}$ and $Sn_{0.6}Sb_{0.4}$ show the same trend with the average power as in the case of $Sn_{0.4}Sb_{0.6}$. The kinetic at four different wavelengths are also fitted with the help of surface Xplorer in both the remaining SnSb alloys.

S.Table IV, V, and VI show the fitting coefficients for $Sn_{0.5}Sb_{0.5}$ at three different average power respectively, while the S.Table VII, VIII, and IX show for $Sn_{0.6}Sb_{0.4}$. The differential reflectance confirms the band filling phenomenon as well as the bleaching effect in the crystals. The kinetic also supports the agreement of band filling as well as the bleaching effect in the systems. Comparing all these crystals at one average power i.e 1 mW, a unique trend can easily be noticeable. Figure 6 shows the appraisal of the 3D pictorial diagram in which the differential reflectance signal with respect to probe wavelength up to a few nanosecond long time is revealed. It shows the comparison of all the SnSb alloys at 1 mW, starting from $Sn_{0.4}Sb_{0.6}$ to $Sn_{0.5}Sb_{0.5}$ and at the end $Sn_{0.6}Sb_{0.4}$. There is a huge difference in differential reflection among the crystals. The $Sn_{0.5}Sb_{0.5}$ has the lowest differential reflectance and the $Sn_{0.4}Sb_{0.6}$ has the highest differential reflectance in magnitude. This similar uneven trend is also present in the case of kinetics among this crystal. For x = 0.5, the magnitude of the kinetic is lowest while for x = 0.4 is more in magnitude which is only due to the compositional change. For an equal number of atoms as in $Sn_{0.5}Sb_{0.5}$, the Sn and Sb are equal, there is low differential reflectance, while any deviation from this increases the differential reflectance. This can be understood in terms of charge carriers. In an SnSb alloy, where Sn and Sb are equal in composition, there is an equal contribution of charge carriers to the system. While any doping in the equilibrium system misbalanced the charge carriers. Any disproportion in the equilibrium SnSb composition increases one type of carriers in some system and another type in others. As in $Sn_{0.6}Sb_{0.4}$, there is more charge carriers induce because of Sn, while in $Sn_{0.4}Sb_{0.6}$, it is because of Sb.

Apart from differential reflectance and kinetic spectra, these SnSb alloys show a unique trend in the fitting parameters of their respective kinetics. A common trend that is present in all the SnSb alloys i.e $t_1$ is almost constant concerning the average power while it increases directly with pump wavelengths for example as shown in figure 5(a) and 5(b). The average power dependency of t1 confirms that the time taken by carriers to bleach is independent of average power, while the directly proportional character represents that more charge carriers accumulated in the lower energy level for a longer time as seen in the differential reflectance signal too. The charge carriers are pumped from the valence band to a higher excited state of 2.61 eV. Thus, upper energy levels like 900 nm, 1100 nm fills first and band filling phenomenon observed in the bands after 1.5 ps due to which lower energy charge carrier persists for a longer time. But, comparing different alloys, the t1 decreases with 'x' as it goes from 0.4 to 0.6. It provides a piece of evidence that time for bleaching charge carriers reduces with the decrease in Sb concentration. While taking into account t2 for SnSb alloys as shown in figure 7, the most stable energy level is 0.95 eV where band filling occurs for a high ps time.

**CONCLUSION**

The previous study of superconducting properties of $Sn_{0.4}Sb_{0.6}$ has been extended to determine the optical properties in this article. The optical studies are further extended to the compositional variation of Sn and Sb in $Sn_xSb_{1-x}$ samples. The optical properties of $Sn_{0.4}Sb_{0.6}$ are thoroughly examined here. The behavior of the sample as a type-II superconductor has been shown using the QD-PPMS (Physical Property Measurement System) which is its topological superconductivity. This topological behaviour is confirmed through DFT as well. The average power-dependent excited state dynamics confirms the presence of band filling and bleaching phenomenon. Thereby, there is a decrease in the refractive index and absorption coefficient in the band filling case, while the real part of the refractive index increase in the bleaching phenomenon. Apart from the average power-dependent phenomenon, a comparison between different alloys confirms the bleaching and band filling as a universal phenomenon in the SnSb alloys. The appraisal also confirms that any doping in the equilibrium system causes an increase in carrier density in the particular alloy. The decay lifetime of the excited carriers helps to predict 0.95 eV of bandgap between band filled states and valence bands too. Thus, this study shows the potential for nonlinear optical applications of topological insulating alloys.

**ACKNOWLEDGMENT**

The director of CSIR-NPL highly supports the work. The authors would like to thanks Dr. J.S Tawale for SEM imaging and Dr. Manika Khanuja for PXRD data. Mr. Prince Sharma likes to thank CSIR-UGC-SRF for the scholarship and AcSIR for enrolling as a Senior research scholar.

**CAPTIONS**

**Figure 1.** Rietveld analysis of the PXRD data and the crystal structure using the parameters in VESTA software. The inset of the figure shows the FESEM image of the crystal at a 2 μm magnification and M-H curve at a particular range of low temperatures using 2-3.7 K using QD-PPMS (Physical Property Measurement System).

**Figure 2.** Comparison of PXRD data among all the crystals of the SnSb alloys showing the peak shift with an increase in Sb concentration and confirms the presence of dual-phase in $Sn_{0.4}Sb_{0.6}$ and $Sn_{0.5}Sb_{0.5}$. Inset shows (a) SEM image of $Sn_{0.5}Sb_{0.5}$ and (b) $Sn_{0.6}Sb_{0.4}$ respectively.

**Figure 3.** The three-dimensional pictorial response of the excited state charge carrier in terms of differential reflectance as the Z-axis. The X-axis and Y-axis represent the probe wavelength and probe delay, respectively.

**Figure 4.** The differential reflectance response of the flake excited by the 2.61 eV laser pulse with the average power of 0.5 mW concerning (a) probe wavelength in the NIR regime and (b) probe decay from few femtoseconds to 5 ns.

**Figure 5(a).** Variation of $t_1$ and $t_2$ excited charge carrier decay lifetime with average power in mW at four different probe wavelengths.

**Figure 5(b).** Bandgap analysis of the crystal from the disparity of $t_1$ and $t_2$ concerning the probe wavelength at different average power.

**Figure 6.** A comparison among all the crystals starting from $Sn_{0.4}Sb_{0.6}$ and end to $Sn_{0.6}Sb_{0.4}$. These three-dimensional pictorial responses of the excited state charge carrier are at 1.0 mW average power. The X-axis and Y-axis represent the probe wavelength and probe delay, respectively. While the Z-axis represents the differential reflectance.

**Figure 7.** Plot of carrier lifetime of all the SnSb alloys and shows the variation of charge carrier decay lifetime with average power in mW at four different probes.

**Figure 1.**

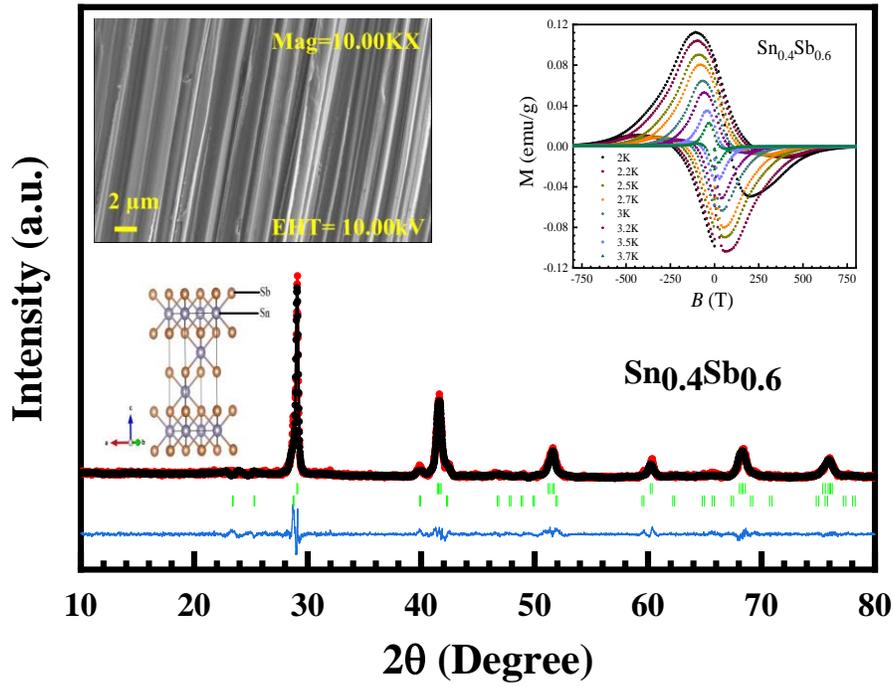

**Figure 2.**

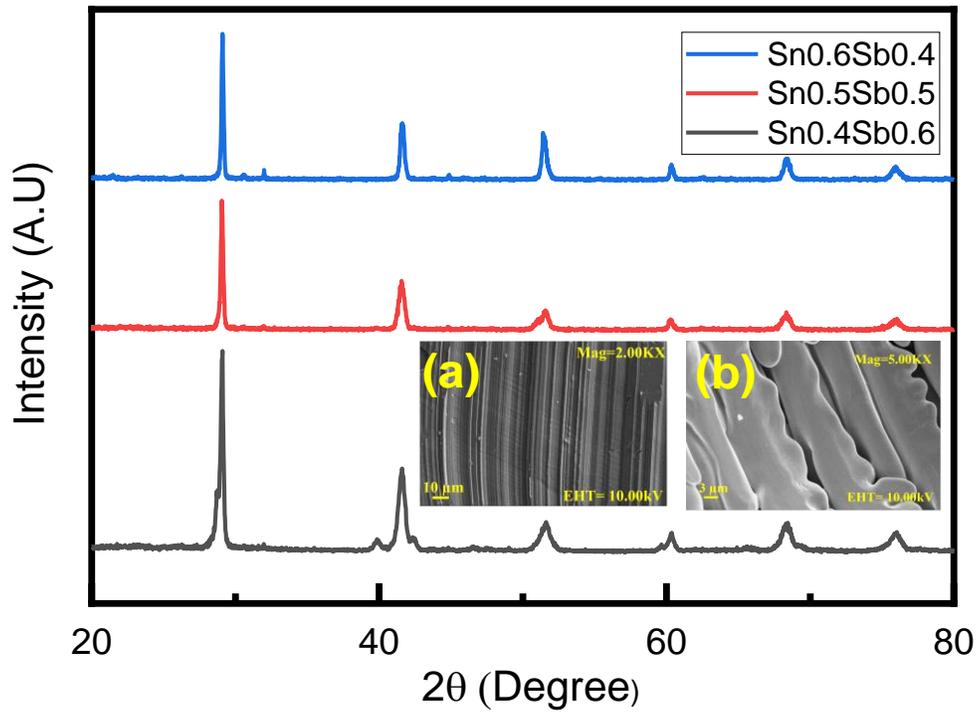

**Figure 3.**

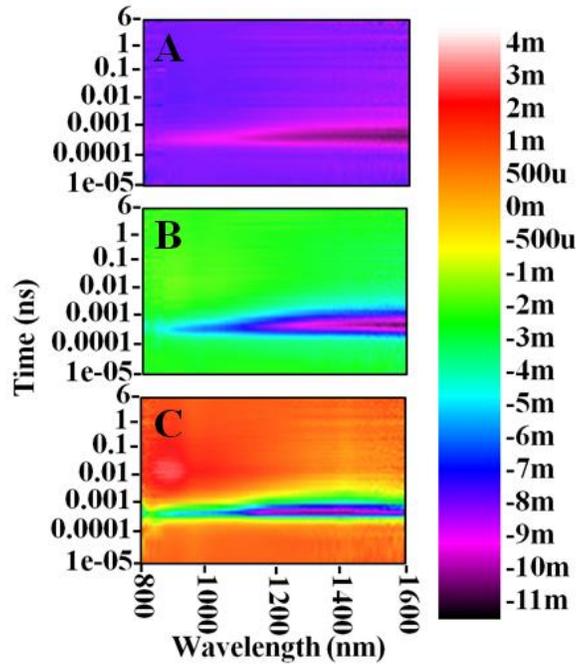

**Figure 4.**

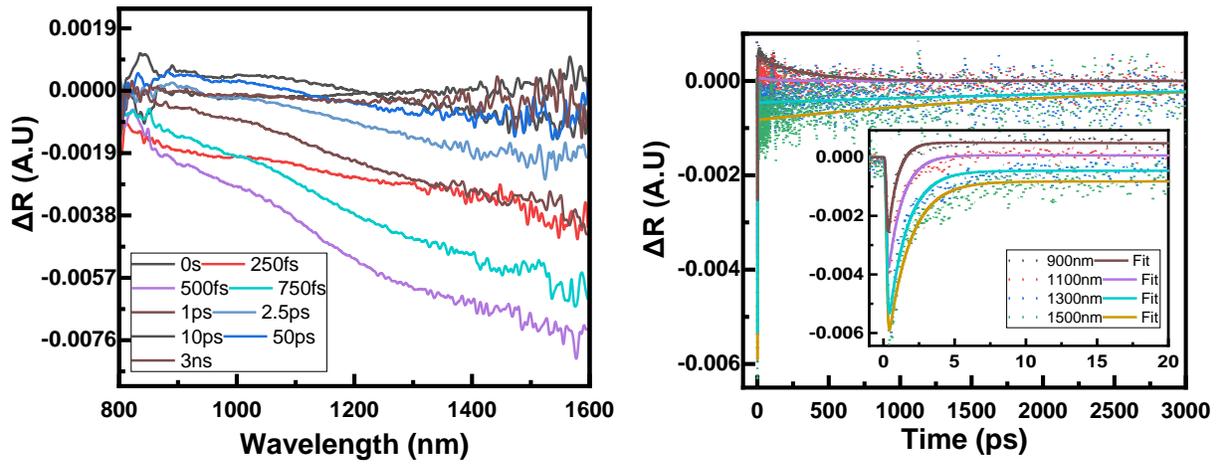

**Figure 5(a).**

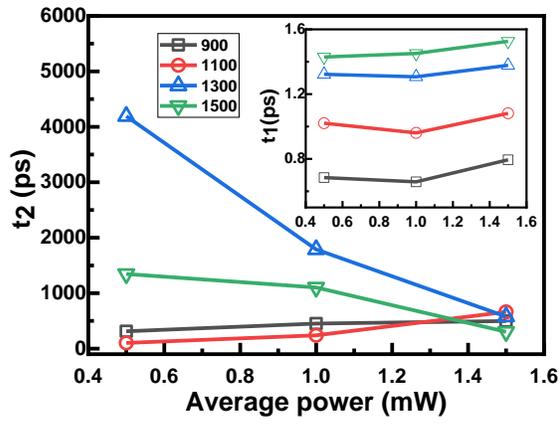

**Figure 5(b).**

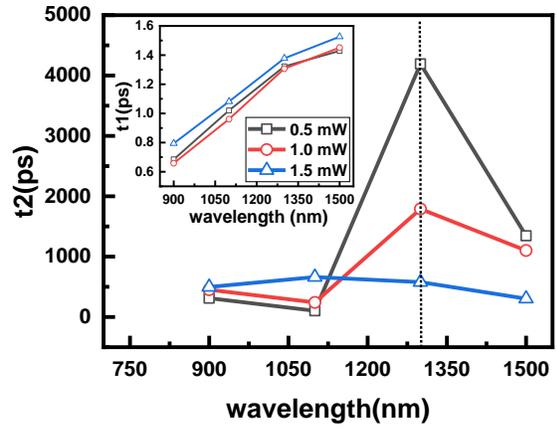

**Figure 6.**

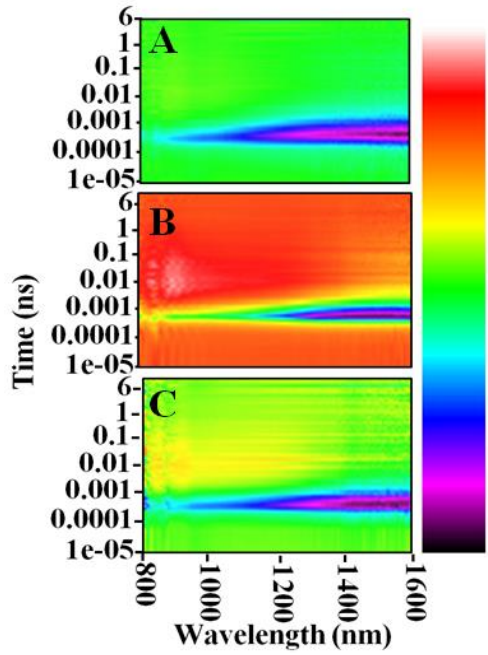

**Figure 7.**

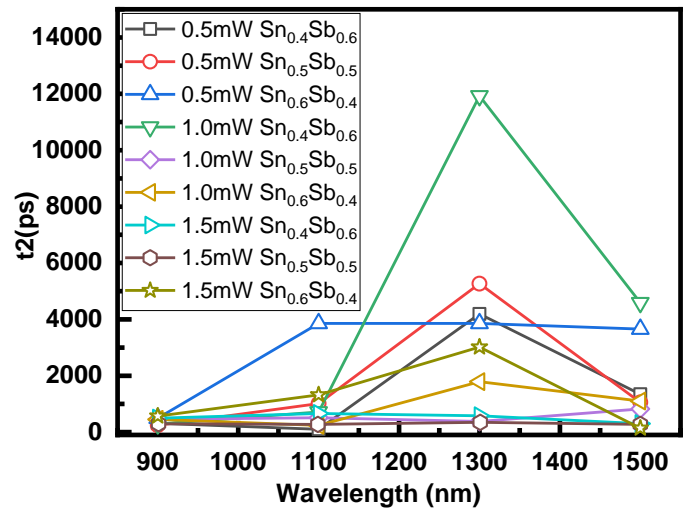

# REFERENCES

1. Bansil A, Lin H, Das T (2016) Colloquium : Topological band theory. Reviews of Modern Physics 88:021004. https://doi.org/10.1103/RevModPhys.88.021004

2. Ando Y, Society P, April R, Ando Y (2013) Topological Insulator Materials. Journal of the Physical Society of Japan 82:1–32. https://doi.org/10.7566/JPSJ.82.102001

3. Fu L, Kane CL, Mele EJ (2007) Topological insulators in three dimensions. Physical Review Letters 98:106803. https://doi.org/10.1103/PhysRevLett.98.106803

4. Hasan MZ, Kane CL (2010) Colloquium: Topological insulators. Reviews of Modern Physics 82:3045–3067. https://doi.org/10.1103/RevModPhys.82.3045

5. Moore JE (2013) Theory of three-dimensional topological insulators. In: Contemporary Concepts of Condensed Matter Science. Elsevier B.V., pp 35–57

6. Yan B, Zhang SC (2012) Topological materials. Reports on Progress in Physics 75:096501. https://doi.org/10.1088/0034-4885/75/9/096501

7. Fu L, Kane CL (2007) Topological insulators with inversion symmetry. Physical Review B - Condensed Matter and Materials Physics 76:1–17. https://doi.org/10.1103/PhysRevB.76.045302

8. Keimer B, Moore JE (2017) The physics of quantum materials. Nature Physics 13:1045–1055. https://doi.org/10.1038/nphys4302

9. Kane CL, Mele EJ (2005) Z2 topological order and the quantum spin hall effect. Physical Review Letters 95:3–6. https://doi.org/10.1103/PhysRevLett.95.146802

10. Yue Z, Wang X, Gu M (2019) Topological Insulator Materials for Advanced Optoelectronic Devices. In: Advanced Topological Insulators. John Wiley & Sons, Inc., Hoboken, NJ, USA, pp 45–70

11. Lu L, Joannopoulos JD, Soljačić M (2014) Topological photonics. Nature Photonics 8:821–829. https://doi.org/10.1038/nphoton.2014.248

12. Qi XL, Zhang SC (2011) Topological insulators and superconductors. Reviews of Modern Physics 83:1057–1110. https://doi.org/10.1103/RevModPhys.83.1057

13. Tang F, Po HC, Vishwanath A, Wan X (2019) Efficient topological materials discovery using symmetry indicators. Nature Physics 15:470–476. https://doi.org/10.1038/s41567-019-0418-7

14. Hsieh D, Qian D, Wray L, et al (2008) A topological Dirac insulator in a quantum spin Hall phase. Nature 452:970–974. https://doi.org/10.1038/nature06843

15. Alicea J (2012) New directions in the pursuit of Majorana fermions in solid state systems. Reports on Progress in Physics 75:076501. https://doi.org/10.1088/0034-4885/75/7/076501

16. Wray LA, Xu SY, Xia Y, et al (2010) Observation of topological order in a superconducting doped topological insulator. Nature Physics 6:855–859. https://doi.org/10.1038/nphys1762

17. Moore J (2009) Topological insulators: The next generation. Nature Physics 5:378–380. https://doi.org/10.1038/nphys1294

18. Sharma P, Kumar M, Awana VPS (2020) Exploration of terahertz from time-resolved ultrafast spectroscopy in single-crystal Bi2Se3 topological insulator. Journal of Materials Science: Materials in Electronics 31:7959–7967. https://doi.org/10.1007/s10854-020-03335-5

19. Sharma P, Sharma MM, Kumar M, Awana VPS (2020) Metal doping in topological insulators- a key for tunable generation of terahertz. Solid State Communications 319:114005. https://doi.org/10.1016/j.ssc.2020.114005

20. Hor YS, Williams AJ, Checkelsky JG, et al (2010) Superconductivity in CuxBi2Se3 and its implications for pairing in the undoped topological insulator. Physical Review Letters 104:057001. https://doi.org/10.1103/PhysRevLett.104.057001

21. Wang Z, Taskin AA, Frölich T, et al (2016) Superconductivity in Tl0.6Bi2Te3 Derived from a Topological Insulator. Chemistry of Materials 28:779–784. https://doi.org/10.1021/acs.chemmater.5b03727

22. Sharma MM, Rani P, Sang L, et al (2020) Superconductivity Below 2.5K in Nb0.25Bi2Se3 Topological Insulator Single Crystal. Journal of Superconductivity and Novel Magnetism 33:565–568. https://doi.org/10.1007/s10948-019-05373-5

23. Yonezawa S (2018) Nematic Superconductivity in Doped Bi2Se3 Topological Superconductors. Condensed Matter 4:2. https://doi.org/10.3390/condmat4010002

24. Han CQ, Li H, Chen WJ, et al (2015) Electronic structure of a superconducting topological insulator Sr-doped Bi2Se3. Applied Physics Letters 107:171602. https://doi.org/10.1063/1.4934590

25. Sharma MM, Sang L, Rani P, et al (2020) Bulk Superconductivity Below 6 K in PdBi2Te3 Topological Single Crystal. Journal of Superconductivity and Novel Magnetism. https://doi.org/10.1007/s10948-019-05417-w

26. Schmetterer C, Polt J, Flandorfer H (2017) The phase equilibria in the Sb-Sn system – Part I: Literature review. Journal of Alloys and Compounds 728:497–505. https://doi.org/10.1016/j.jallcom.2017.08.215

27. Schmetterer C, Polt J, Flandorfer H (2018) The phase equilibria in the Sb-Sn system – Part II: Experimental results. Journal of Alloys and Compounds 743:523–536. https://doi.org/10.1016/j.jallcom.2017.11.367

28. Liu B, Wu J, Cui Y, et al (2018) Superconductivity in SnSb with a natural superlattice structure. Superconductor Science and Technology 31:125011. https://doi.org/10.1088/1361-6668/aae6fe

29. Liu B, Xiao C, Zhu Q, et al (2019) Superconducting phase diagram and nontrivial band

topology of structurally modulated Sn1-xSbx. Physical Review Materials 3:084603. https://doi.org/10.1103/PhysRevMaterials.3.084603

30. M.M. Sharma, Kapil Kumar, Lina Sang XLW and VPSA, Sharma MM, Kumar K, et al (2020) Type-II Superconductivity below 4K in Sn0.4Sb0.6. Journal of Alloys and Compounds 844:156140. https://doi.org/10.1016/j.jallcom.2020.156140

31. Seo DM, Lee JH, Lee S, et al (2019) Ultrafast Excitonic Behavior in Two-Dimensional Metal-Semiconductor Heterostructure. ACS Photonics 6:1379–1386. https://doi.org/10.1021/acsphotonics.9b00399

32. Gupta V, Bharti V, Kumar M, et al (2015) Polymer-Polymer Förster Resonance Energy Transfer Significantly Boosts the Power Conversion Efficiency of Bulk-Heterojunction Solar Cells. Advanced Materials 27:4398–4404. https://doi.org/10.1002/adma.201501275

33. Hinuma Y, Pizzi G, Kumagai Y, et al (2017) Band structure diagram paths based on crystallography. Computational Materials Science 128:140–184. https://doi.org/10.1016/j.commatsci.2016.10.015

34. Berera R, van Grondelle R, Kennis JTM (2009) Ultrafast transient absorption spectroscopy: Principles and application to photosynthetic systems. Photosynthesis Research 101:105–118

35. Sun D, Lai JW, Ma JC, et al (2017) Review of ultrafast spectroscopy studies of valley carrier dynamics in two-dimensional semiconducting transition metal dichalcogenides. Chinese Physics B 26:037801. https://doi.org/10.1088/1674-1056/26/3/037801

36. Tian L, Di Mario L, Zannier V, et al (2016) Ultrafast carrier dynamics, band-gap renormalization, and optical properties of ZnSe nanowires. Physical Review B 94:165442. https://doi.org/10.1103/PhysRevB.94.165442

37. Williams KW, Monahan NR, Koleske DD, et al (2016) Ultrafast and band-selective Auger recombination in InGaN quantum wells. Applied Physics Letters 108:141105. https://doi.org/10.1063/1.4945669

38. Hsu CC, Lin JH, Chen YS, et al (2008) Ultrafast carrier dynamics of InGaAsN and InGaAs single quantum wells. Journal of Physics D: Applied Physics 41:085107. https://doi.org/10.1088/0022-3727/41/8/085107

39. Braun L, Mussler G, Hruban A, et al (2016) Ultrafast photocurrents at the surface of the three-dimensional topological insulator Bi2Se3. Nature Communications 7:13259. https://doi.org/10.1038/ncomms13259

40. Foggi P, Bussotti L, Neuwahl FVR (2001) Photophysical and photochemical applications of femtosecond time-resolved transient absorption spectroscopy. International Journal of Photoenergy 3:103–109. https://doi.org/10.1155/S1110662X01000125

41. Ruckebusch C, Sliwa M, Pernot P, et al (2012) Comprehensive data analysis of femtosecond transient absorption spectra: A review. Journal of Photochemistry and Photobiology C: Photochemistry Reviews 13:1–27

# Excited-state dynamics of structurally characterized crystal of $Sn_xSb_{1-x}$

Prince Sharma[1,2], M.M. Sharma[1,2], Kapil Kumar[1,2], Mahesh Kumar[1,2], V.P.S. Awana[1,2*]

[1]CSIR-National Physical Laboratory, Dr. K.S. Krishnan Marg, New Delhi-110012, India
[2]Academy of Scientific and Innovative Research (AcSIR), Ghaziabad, 201002, India

**SUPPLEMENTARY INFORMATION**

**Figure 1. Optimized heat treatment used for preparing the SnSb alloys**

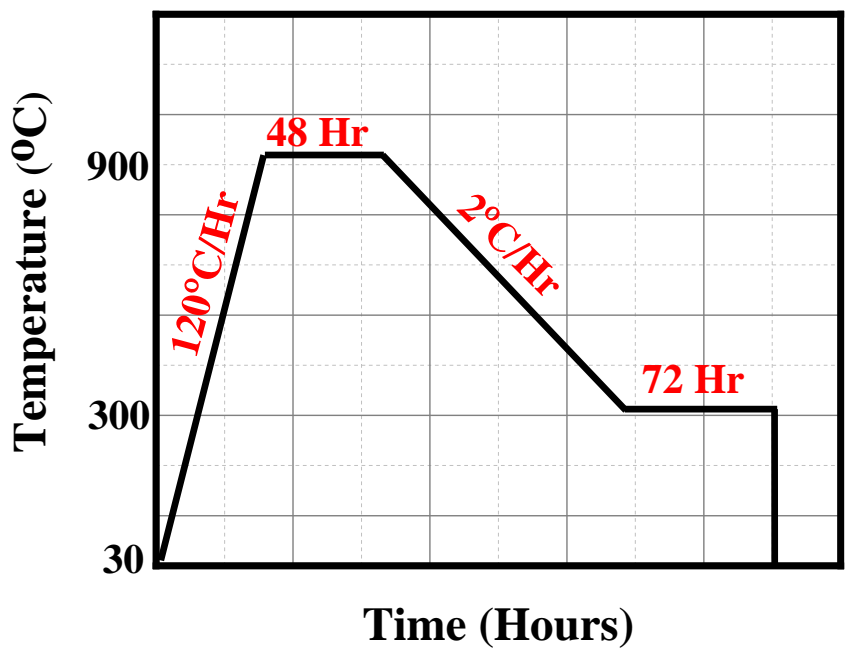

**Figure 2 (a).** The calculated band structure plots along the Γ-H0-L-T-Γ direction of the Brillouin zone without SOC.

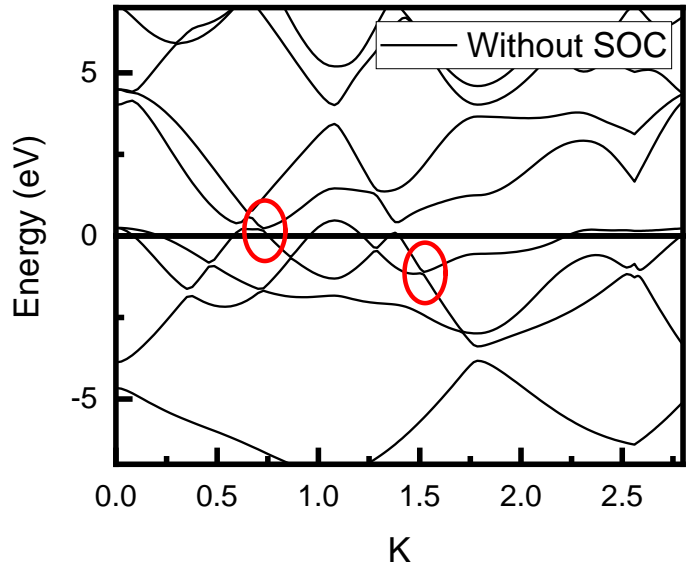

**Figure 2 (b).** The calculated band structure plots along the Γ-H0-L-T-Γ direction of the Brillouin zone with SOC.

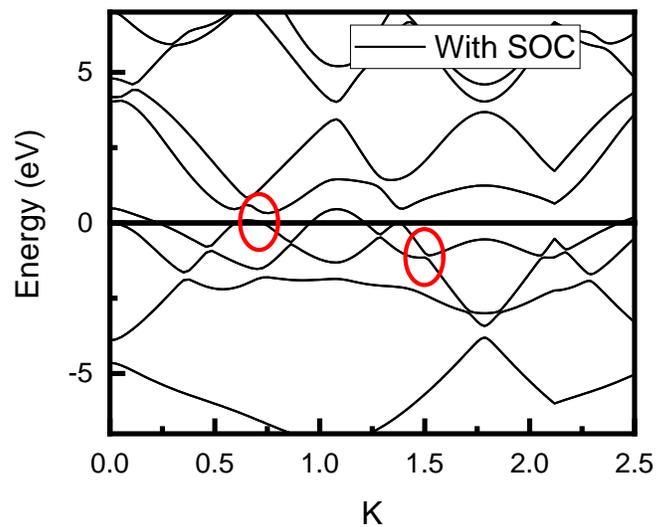

**Figure 3. The differential reflectance response and kinetic decay profile of the flake $Sn_{0.4}Sb_{0.6}$ at an average power of 1.0 mW**

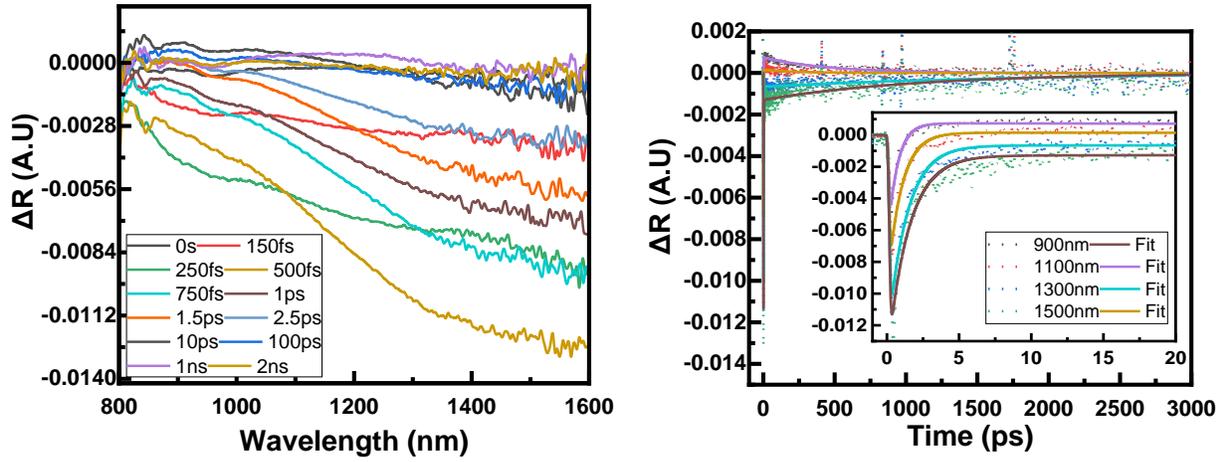

**Figure 4. The differential reflectance response and kinetic decay profile of the flake $Sn_{0.4}Sb_{0.6}$ at an average power of 1.5 mW**

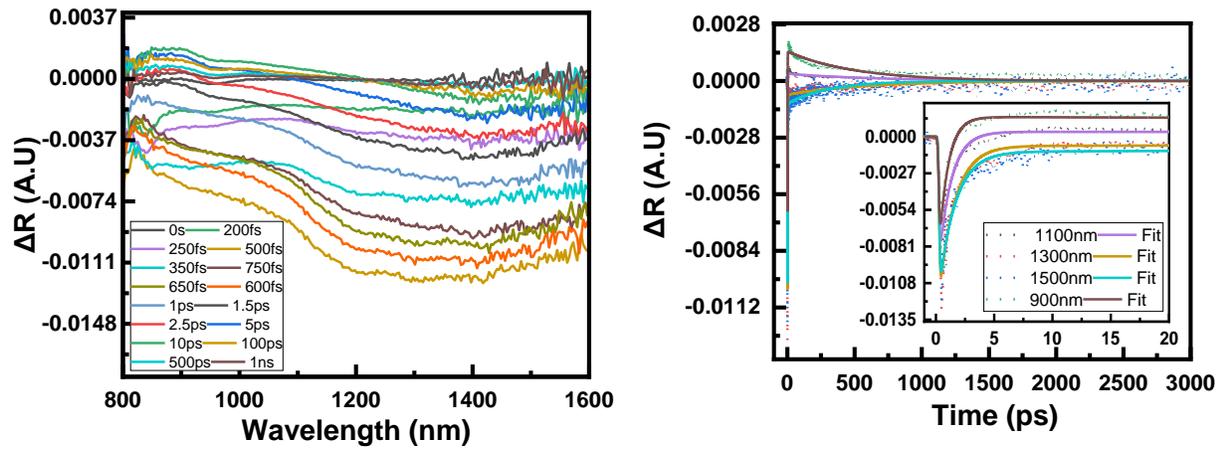

**Figure 5.** The three-dimensional pictorial response of the excited state charge carrier of $Sn_{0.5}Sb_{0.5}$ in terms of differential reflectance with respect to probe wavelength and probe delay, respectively.

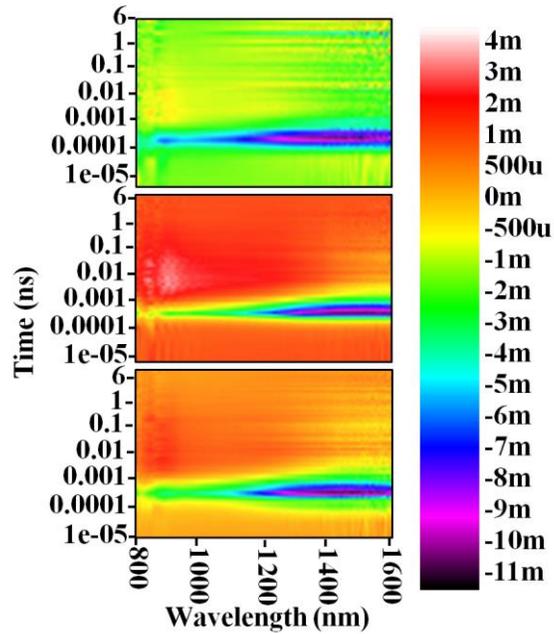

**Figure 6.** The three-dimensional pictorial response of the excited state charge carrier of $Sn_{0.6}Sb_{0.4}$ in terms of differential reflectance with respect to probe wavelength and probe delay, respectively.

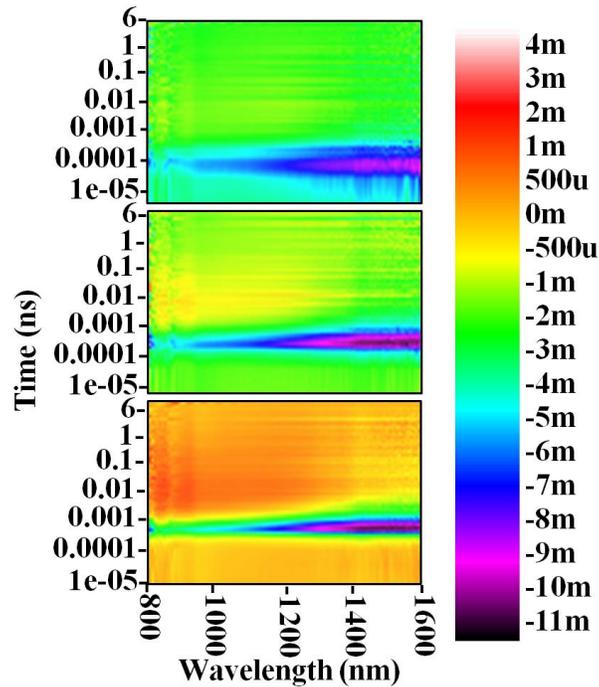

**Figure 7.** The differential reflectance response and kinetic decay profile of the flake $Sn_{0.5}Sb_{0.5}$ at an average power of 0.5 mW

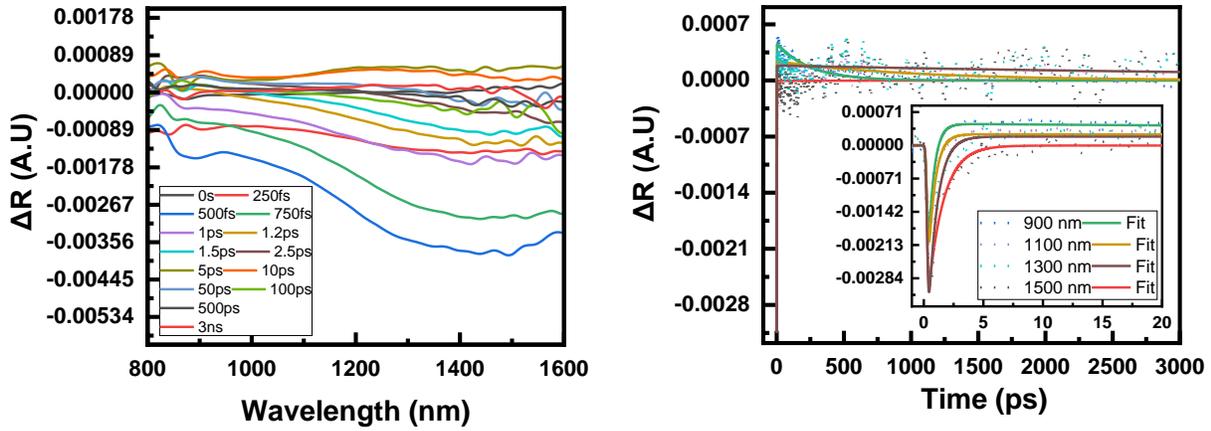

**Figure 8.** The differential reflectance response and kinetic decay profile of the flake $Sn_{0.5}Sb_{0.5}$ at an average power of 1.0 mW

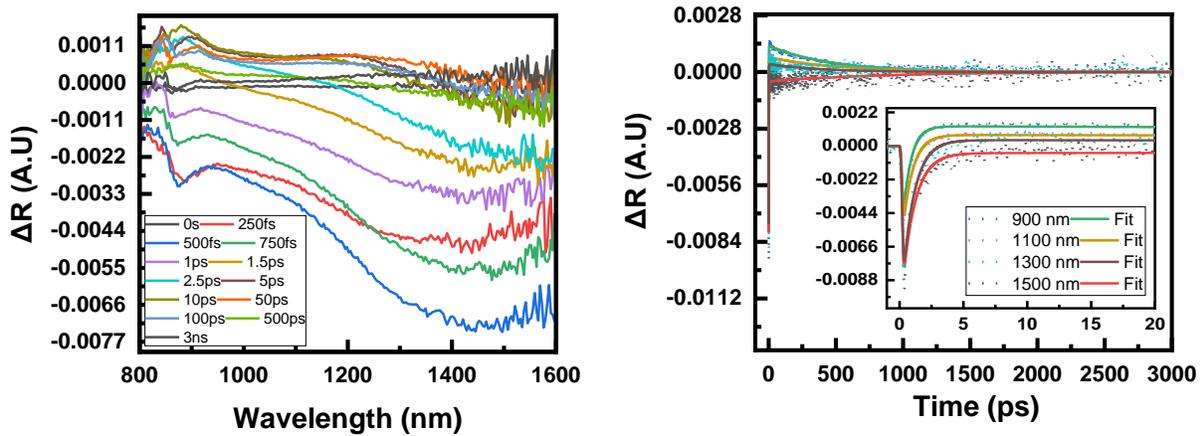

**Figure 9. The differential reflectance response and kinetic decay profile of the flake $Sn_{0.5}Sb_{0.5}$ at an average power of 1.5 mW**

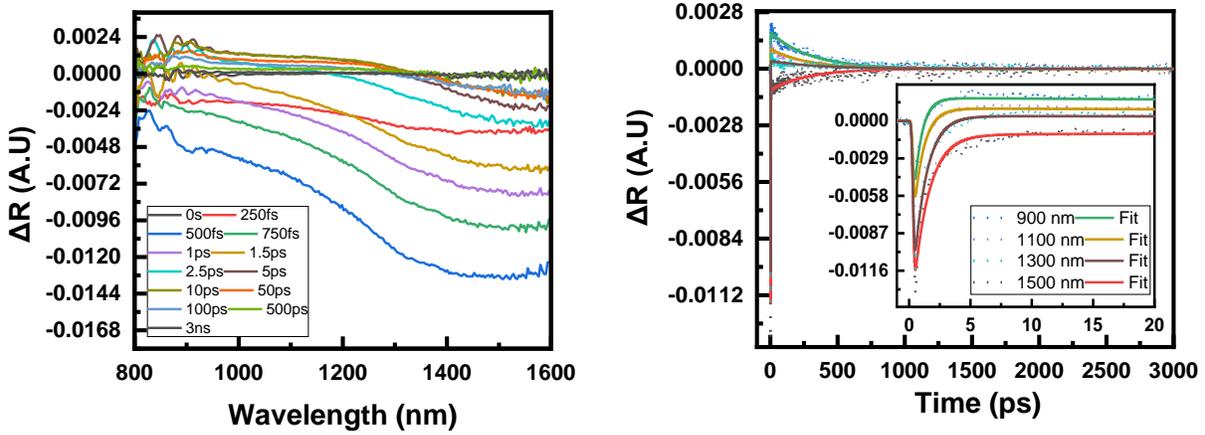

**Figure 10. The differential reflectance response and kinetic decay profile of the flake $Sn_{0.6}Sb_{0.4}$ at an average power of 0.5 mW**

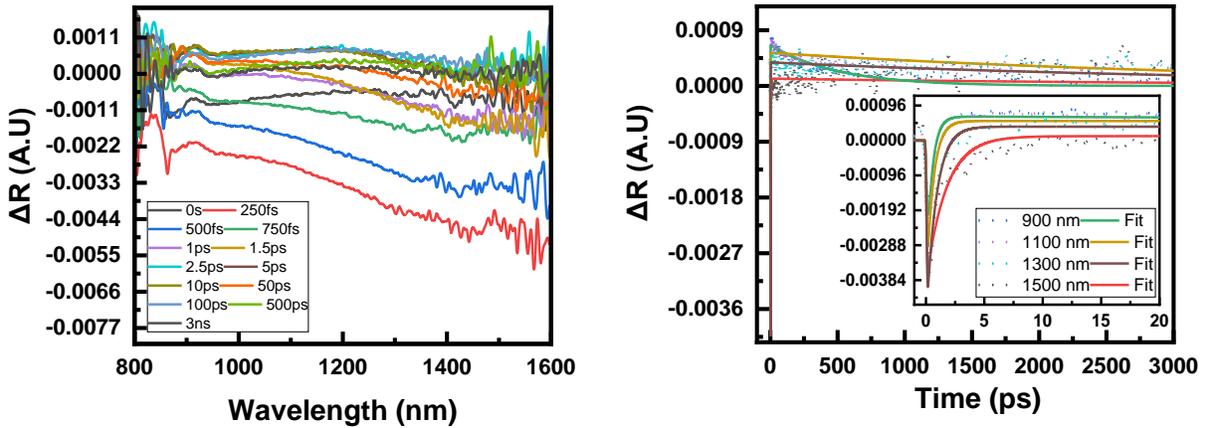

**Figure 11.** The differential reflectance response and kinetic decay profile of the flake $Sn_{0.6}Sb_{0.4}$ at an average power of 1.0 mW

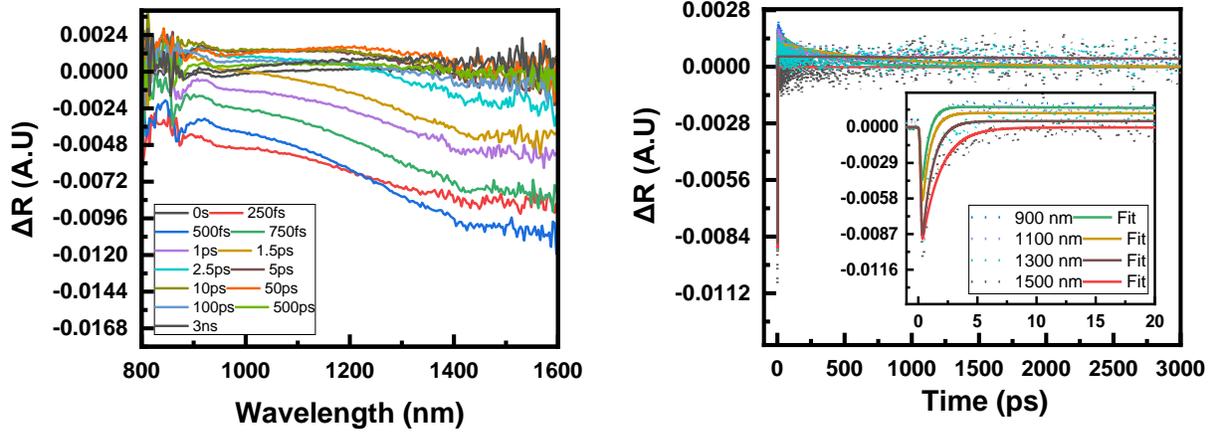

**Figure 12.** The differential reflectance response and kinetic decay profile of the flake $Sn_{0.6}Sb_{0.4}$ at an average power of 1.5 mW

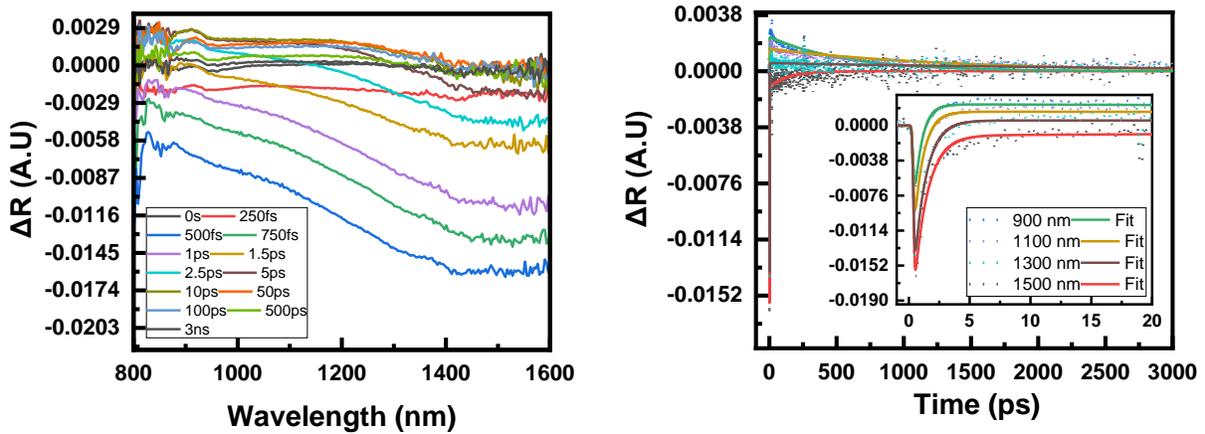

**Table I.** The excited charge carriers decay lifetime and the amplitude of decay of $Sn_{0.4}Sb_{0.6}$ when excited by the 0.5 mW average power. The fitting of the kinetic decay with the help of surface Xplorer gives the parameters.

| wavelength | t1(ps) | A1(%) | t2(ps) | A2(%) |
|---|---|---|---|---|
| 900 | 0.68 | 88.16 | 312.25 | 11.83 |
| 1100 | 1.02 | 98.40 | 102.21 | 1.59 |
| 1300 | 1.32 | 92.59 | 4192.69 | 7.40 |
| 1500 | 1.42 | 89.26 | 1345.21 | 10.73 |

**Table II.** The excited charge carriers decay lifetime and the amplitude of decay of $Sn_{0.4}Sb_{0.6}$ when excited by the 1.0 mW average power.

| wavelength | t1(ps) | A1(%) | t2(ps) | A2(%) |
|---|---|---|---|---|
| 900 | 0.65 | 89.74 | 451.33 | 10.25 |
| 1100 | 0.96 | 98.26 | 240.46 | 1.73 |
| 1300 | 1.30 | 94.38 | 1789.07 | 5.61 |
| 1500 | 1.45 | 90 | 1101.98 | 10.00 |

**Table III.** The excited charge carriers decay lifetime and the amplitude of decay of $Sn_{0.4}Sb_{0.6}$ when excited by the 1.5 mW average power.

| wavelength | t1(ps) | A1(%) | t2(ps) | A2(%) |
|---|---|---|---|---|
| 900 | 0.79 | 87.22 | 496.63 | 12.77 |
| 1100 | 1.08 | 96.31 | 661.06 | 3.68 |
| 1300 | 1.37 | 94.22 | 578.43 | 5.77 |
| 1500 | 1.41 | 90.07 | 302.45 | 9.92 |

**Table IV. The excited charge carriers decay lifetime and the amplitude of decay of $Sn_{0.5}Sb_{0.5}$ when excited by the 0.5 mW average power.**

| wavelength | t1(ps) | A1(%) | t2(ps) | A2(%) |
|---|---|---|---|---|
| 900 | 0.48 | 88.69 | 243.75 | 11.30 |
| 1100 | 0.57 | 93.47 | 1013.59 | 6.52 |
| 1300 | 0.77 | 95.93 | 5268.37 | 4.06 |
| 1500 | 1.26 | 99.84 | 1055.92 | 0.15 |

**Table V. The excited charge carriers decay lifetime and the amplitude of decay of $Sn_{0.5}Sb_{0.5}$ when excited by the 1.0 mW average power.**

| wavelength | t1(ps) | A1(%) | t2(ps) | A2(%) |
|---|---|---|---|---|
| 900 | 0.62 | 84.08 | 455.47 | 15.91 |
| 1100 | 0.76 | 90.30 | 511.44 | 9.69 |
| 1300 | 0.92 | 96.16 | 379.69 | 3.83 |
| 1500 | 0.98 | 95.02 | 823.29 | 4.97 |

**Table VI. The excited charge carriers decay lifetime and the amplitude of decay of $Sn_{0.5}Sb_{0.5}$ when excited by the 1.5 mW average power.**

| wavelength | t1(ps) | A1(%) | t2(ps) | A2(%) |
|---|---|---|---|---|
| 900 | 0.63 | 83.27 | 297.21 | 16.72 |
| 1100 | 0.82 | 90.16 | 273.74 | 9.83 |
| 1300 | 1.04 | 97.11 | 346.61 | 2.88 |
| 1500 | 1.25 | 92.23 | 275.54 | 7.76 |

**Table VII. The excited charge carriers decay lifetime and the amplitude of decay of $Sn_{0.6}Sb_{0.4}$ when excited by the 0.5 mW average power.**

| wavelength | t1(ps) | A1(%) | t2(ps) | A2(%) |
|---|---|---|---|---|
| 900 | 0.47 | 86.30 | 496.07 | 13.69 |
| 1100 | 0.59 | 89.05 | 3858.17 | 10.94 |
| 1300 | 0.85 | 93.13 | 3858.38 | 6.86 |
| 1500 | 1.68 | 97.09 | 3657.00 | 2.90 |

**Table VIII. The excited charge carriers decay lifetime and the amplitude of decay of $Sn_{0.6}Sb_{0.4}$ when excited by the 1.0 mW average power.**

| wavelength | t1(ps) | A1(%) | t2(ps) | A2(%) |
|---|---|---|---|---|
| 900 | 0.57 | 83.33 | 287.48 | 16.66 |
| 1100 | 0.68 | 88.92 | 716.00 | 11.07 |
| 1300 | 0.89 | 95.72 | 11927.60 | 4.27 |
| 1500 | 1.55 | 99.85 | 4592.28 | 0.14 |

**Table IX. The excited charge carriers decay lifetime and the amplitude of decay of $Sn_{0.6}Sb_{0.4}$ when excited by the 1.5 mW average power.**

| wavelength | t1(ps) | A1(%) | t2(ps) | A2(%) |
|---|---|---|---|---|
| 900 | 0.64 | 84.19 | 568.71 | 15.80 |
| 1100 | 0.77 | 90.22 | 1320.54 | 9.77 |
| 1300 | 0.98 | 96.93 | 3016.52 | 3.06 |
| 1500 | 1.11 | 94.21 | 137.81 | 5.78 |